\documentstyle[aps]{revtex}
\input{psfig.sty}
\newcommand{\equ}[1]{~Eq.~(\ref{#1})}

\newcommand{\sla}{\raise.15ex\hbox{$/$}\kern -.8em} 
 
\newcommand{\half}{\frac 1 2}

\newcommand{\cb}{{\bar c}}

\renewcommand{\L}{\Lambda}
\newcommand{\LQCD}{\L_{\overline{MS}}}

\newcommand{\m}{\mu}\newcommand{\n}{\nu}
\newcommand{\mn}{{\mu\nu}}
\newcommand{\eps}{\varepsilon}
\newcommand{\bra}{\langle}\newcommand{\ket}{\rangle}
\newcommand{\be}[1]{\begin{equation}\label{#1}} 
\newcommand{\ee}{\end{equation}}
\newcommand{\ba}[1]{\begin{eqnarray}\label{#1}} 
\newcommand{\ea}{\end{eqnarray}}

\preprint{NYU-TH/99/08/30}
\draft
\title{Mass Generation in Continuum $SU(2)$ Gauge Theory in  Covariant
Abelian Gauges}   
\author{Martin Schaden}
\address{New York University, Physics Department,
 4 Washington Place, New York, New York 10003}
\date{\today}
\twocolumn
\begin{document}
\maketitle
\begin{abstract}
The local action of an $SU(2)$ gauge theory 
in general covariant Abelian gauges and the associated
equivariant BRST symmetry that guarantees the perturbative
renormalizability  of the model are given. I show that a global SL(2,R)
symmetry of the model is spontaneously broken by ghost-antighost
condensation at arbitrarily small coupling and leads 
to propagators that are finite at Euclidean momenta for all elementary
fields except the Abelian ``photon''.  The Goldstone states form a
BRST-quartet. The mechanism eliminates the non-abelian
infrared divergences in the perturbative high-temperature 
expansion of the free energy.
\end{abstract}   

\noindent The existence of a
quark gluon plasma phase in which quarks  and gluons are weakly
interacting degrees of freedom at temperatures $T\gtrsim\Lambda$   
is suggested by the renormalization group flow of the effective QCD
coupling constant and the analogy with the plasma phase of QED. Lattice
simulations indicate that this phase 
transition is first order and for pure $SU(2)$ occurs at $T^2_c\sim 2.1\times 
{\rm ~string~tension}$\cite{De96}. Although QCD is asymptotically free, 
the perturbative analysis 
of the high-temperature phase is plagued by infrared (IR)
divergences\cite{Ka89}. The best one can presently achieve
perturbatively at high temperatures is a resummation of the 
infrared-safe contributions\cite{Pi89}. The
situation is somewhat embarrassing, since one naively might hope 
that an asymptotically free theory allows for an accurate perturbative
description of the high temperature phase. 
The IR-problem encountered in the perturbative high
temperature expansion in fact is part of the more general problem of defining
a non-abelian gauge-fixed theory on a compact Euclidean space-time
without boundaries, such as a hypertorus. It was shown\cite{Ba98} 
that normalizable zero-modes of the ghosts cause the partition 
function to vanish in conventional covariant gauges. An  
equivariant BRST construction was used to eliminate these ghost
zero-modes associated with global gauge 
invariance at the expense of a non-local quartic ghost interaction.
It was seen that this interaction leads to  ghost-antighost 
condensation at arbitrarily small
coupling\cite{Ba98,Sc98}. Subsequently, the
equivariant gauge-fixing procedure was successfully
used to reduce the structure group of an $SU(2)$ lattice gauge theory (LGT)
to a physically equivalent Abelian LGT with a $U(1)$ structure
group\cite{Sc99}. The equivariant BRST symmetry of the partially
gauge-fixed LGT was proven to be valid also non-perturbatively and the 
associated quartic 
ghost interaction leads to ghost-antighost condensation in this case too.
The starting point of this investigation is a  transcription 
of this partially gauge-fixed SU(2)-LGT to the continuum using the
equivariant BRST algebra. A partially gauge-fixed
LGT together with a corresponding critical continuum model that are defined 
by the same BRST algebra is perhaps of some interest.

We will see that a certain SL(2,R) symmetry associated with this gauge fixing 
is spontaneously broken by ghost-antighost condensation at arbitrarily
small coupling in the continuum model and leads to propagators
that are regular at Euclidean momenta for all fields 
except the Abelian ``photon''. The corresponding massless Goldstone states 
form a BRST-quartet and do not contribute to physical quantities such as the 
free energy.  Screening ``masses''
in a certain sense thus arise naturally in an $SU(2)$ gauge theory 
in covariant Abelian gauges. They cure the IR-problem of the
perturbative sceleton expansion for the free energy.  

The critical continuum action of the lattice model\cite{Sc99} in Euclidean
space is uniquely specified by the BRST algebra, the field content and power
counting.  Decomposing the
non-abelian SU(2) connection $\vec A_\mu=(W_\mu^1,W_\mu^2, A_\mu)$ in
terms of two real vector-bosons (or one complex one) and a
U(1)-connection $A_\mu$ (the ``photon'' of the model), the loop
expansion is defined by the Lagrangian, 
\be{L}
{\cal L}={\cal L}_{\rm inv.}+{\cal L}_{\rm AG}+{\cal L}_{\rm aGF}\ .
\ee
Here ${\cal L}_{\rm inv.}$ is the usual $SU(2)$-invariant Lagrangian
written in terms of the vector bosons and the
photon\footnote{Latin
indices take values in $\{1,2\}$ only, Einstein's summation
convention applies and $\eps^{12}=-\eps^{21}=1$, vanishing 
otherwise. All results are in the $\overline{MS}$ renormalization  
scheme.}, 
\be{Linv}
{\cal L}_{\rm inv.}={\cal L}_{\rm matter}+ {\frac 1 4} (G_\mn G_\mn+ G_\mn^a
G_\mn^a)\ ,
\ee 
with
\ba{G}
G_\mn&=&\partial_\m A_\n-\partial_\n A_\m -g\eps^{ab} W_\m^a W_\n^b\nonumber\\
G_\mn^a&=&D^{ab}_\m W_\n^b- D^{ab}_\n W_\m^b\nonumber\\
&=&\partial_\m W_\n^a
-\partial_\n W_\m^a+g\eps^{ab}(A_\m W_\n^b-A_\n W_\m^b)\ .
\ea
${\cal L}_{\rm AG}$ partially gauge-fixes to the maximal
Abelian subgroup $U(1)$ of $SU(2)$ in a covariant manner,  
\be{MAG}
{\cal L}_{AG}=\frac{F^a F^a}{2\alpha}-\cb^a M^{ab} c^b -g^2
\frac{\alpha}{2} (\cb^a\eps^{ab} c^b)^2\ ,
\ee
with 
\ba{defs}
F^a&=& D_\m^{ab} W_\m^b=\partial_\m W_\m^a+ g A_\m\eps^{ab} W_\m^b\nonumber\\
M^{ab}&=& D_\m^{ac} D_\m^{cb} + g^2\eps^{ac}\eps^{bd} W_\m^c
W_\m^d\ .
\ea 
Note that $L_{U(1)}=L_{\rm inv.}+{\cal L}_{\rm AG}$ is invariant
under $U(1)$-gauge transformations {\it and} under an on-shell 
BRST symmetry $s$ and anti-BRST symmetry $\bar s$, whose action on the
fields is
\begin{displaymath}
\begin{array}{rclcrcl}
s A_\m&=&g\eps^{ab}c^aW_\m^b&& \bar sA_\m&=&g\eps^{ab}\cb^aW_\m^b\\
s W_\m^a&=& D_\m^{ab} c^b&& \bar s W_\m^a&=& D_\m^{ab} \cb^b\\
s c^a&=&0&&\bar s \cb^a&=&0\\
s \cb^a&=&F^a/\alpha&&\bar s c^a&=&-F^a/\alpha\ ,
\end{array}
\end{displaymath}
\vspace{-1.0cm}\be{brs}\ee
with an obvious extension to include matter fields. 
Contrary to most other proposals for mass generation\cite{Cu76,Bl96}
the BRST algebra\equ{brs} closes on-shell 
on the set of $U(1)$-invariant functionals: on functionals that 
depend only on $W,A,c$ and the matter fields, $s^2$ for instance 
effects an infinitesimal U(1)-transformation with the parameter ${\frac g
2}\eps^{ab}c^a c^b$. The algebra \equ{brs} thus defines an
equivariant cohomology and ensures the perturbative renormalizability
and unitarity \cite{Sc99,Be76} of the model. 
Note that the physical sector comprises states created by composite
operators of $A,W$ and the matter fields in the equivariant cohomology
of $s$ (or $\bar s$). They are BRST closed {\it and} $U(1)$-invariant. 

The corresponding equivariant (anti-) BRST symmetry
of the LGT is valid also non-perturbatively and it was 
shown\cite{Sc99} that expectation values of physical observables of the
U(1)-LGT are the same as those of the original
$SU(2)$-LGT for any $\alpha>0$. Note that formally setting $\alpha=0$ and 
solving the constraint $F^a=0$ as in\cite{Re98} is {\it not} 
the same as taking the limit $\alpha\rightarrow 0$. The
reason is inherently non-perturbative and nicely exhibited by the lattice
calculation\cite{Sc99}: without the quartic ghost interaction,
Gribov copies of a configuration conspire to give {\it vanishing} 
expectation values for all physical observables. No matter how small,
the quartic ghost interaction is required to have a
normalizable partition function and expectation values of physical
observables that are identical with those of the original SU(2)-LGT.

${\cal L}_{\rm aGF}$ in\equ{L} has been added ``by hand'' to fix the
remaining $U(1)$ gauge invariance and define the perturbative 
series of the continuum model unambiguously. I will assume a 
conventional covariant gauge-fixing term,
\be{aGF}
L_{\rm aGF}=\frac{(\partial_\m A_\m)^2}{2\xi}\ .
\ee 
However, none of the following conclusions  depend on the
gauge-fixing of the Abelian subgroup -- they in particular do not 
depend on $\xi$. 

The Lagrangian \equ{L} also is invariant under a global bosonic
SL(2,R) symmetry generated by  
\be{SL2}
\Pi^+=\int c^a(x)\frac{\delta}{\delta \cb^a(x)}\ \ ,\ \  
\Pi^-=\int \cb^a(x)\frac{\delta}{\delta c^a(x)}\ ,
\ee
and the ghost number $\Pi=[\Pi^+,\Pi^-]$. This SL(2,R) symmetry is 
is also realized in the lattice regularized model\cite{Sc99} and
not anomalous. The conserved currents corresponding to $\Pi^\pm$ 
are U(1)-invariant and BRST, respectively anti-BRST exact,
\be{currents}
j^+_\m=c^a D^{ab}_\m c^b=s c^aW^a_\m\ ,\ \ j_\m^-=\cb^a
D^{ab}_\m\cb^b=\bar s \cb^aW_\m^a\ . 
\ee
I will argue that this global $SL(2,R)$ symmetry of the
model is spontaneously broken to the noncompact abelian subgroup
generated  by the ghost number $\Pi$. Because the
currents\equ{currents} are (anti)-BRST exact, a
spontaneously broken SL(2,R) symmetry is accompanied by a BRST-quartet
of massless Goldstone states with ghost numbers $2,1,-1$ and $-2$. They are
U(1)-invariant $c-c$, $c-W$, $\cb-W$ and $\cb-\cb$ bound states. 
It is important to note that BRST quartets do not
contribute to physical quantities\cite{Ku78} such as the
free energy\footnote{This is analogous to the decoupling of the
Goldstone quartets of the weak interaction  in $R_\xi$
gauges\cite{Ku78}.}. The spontaneous symmetry
breaking in this sense is similar to a (dynamical) Higgs mechanism in
the adjoint. 

An order parameter for the spontaneous breaking of the
SL(2,R) symmetry is
\be{opar}
\bra\cb^a\eps^{ab}c^b\ket=\half\bra\Pi^-
(c^a\eps^{ab}c^b)\ket=-\half\bra\Pi^+ (\cb^a\eps^{ab} \cb^b)\ket\ . 
\ee
To perturbatively investigate the consequences of
$\bra\cb^a\eps^{ab}c^b\ket\neq 0$, the quartic ghost
interaction in\equ{MAG} is linearized using an auxiliary scalar
field $\rho(x)$ of canonical dimension two. Adding the quadratic term 
\be{Stratanovic}
{\cal L}_{\rm aux}=\frac{1}{2 g^2} (\rho -g^2\lambda \cb^a\eps^{ab} c^b)^2
\ee
to the Lagrangian of\equ{L}, the tree level quartic ghost interaction
 vanishes at $\lambda^2=\alpha $
and is then formally of $O(g^4)$, proportional to the difference 
$Z^2_\lambda-Z_\alpha$ of the renormalization constants of the two 
couplings\footnote{The discrete symmetry $ c^a\rightarrow \cb^a,\
\cb^a\rightarrow -c^a,\ \rho\rightarrow -\rho$ relating $s$ and $\bar
s$ also ensures that $\rho$ only mixes with $\cb^a\eps^{ab}c^b$.}.

We shall see that the perturbative expansion
about a {\it non-trivial} solution  $\langle\rho\rangle=v\neq 0$ to the 
gap equation
\be{gap}
\frac{v}{g^2}=\sqrt{\alpha}\left.\bra c^a(x)\eps^{ab}{\bar
c}^b(x)\ket \right|_{<\rho>=v}\ ,
\ee
is much better behaved in the infrared. Note that \equ{gap} is
$U(1)$-invariant and therefore does not depend on the $U(1)$
gauge-fixing\equ{aGF}.  Let us for the moment assume that
a unique non-trivial solution to\equ{gap} exists in some gauge
$\alpha$;  we return to this conjecture below.  The consequences for the
IR-behavior of the model are dramatic. Defining the
quantum part $\sigma(x)$ of the auxiliary scalar $\rho$ by
\be{decomp}
\rho(x)=v+\sigma(x) \ \ {\rm with}\ \ \bra\sigma\ket=0\ ,
\ee
the momentum representation of the Euclidean ghost propagator at tree
level becomes 
\be{ghost}
\bra c^a \cb^b\ket_p
=\frac{p^2\delta^{ab}+\sqrt{\alpha}v\eps^{ab}}{p^4+\alpha v^2}\ .\\
\ee
Feynman's parameterization of this
propagator allows an
evaluation of loop integrals using dimensional regularization  that is
only slightly more complicated than usual. More importantly, 
the ghost propagator is regular at 
Euclidean momenta  when $v\neq 0$. Its complex conjugate poles at
$p^2=\pm i\sqrt{\alpha v^2}$ can furthermore 
not be interpreted as due to asymptotic ghost states\cite{St99}.

When $v\neq 0$, the $W$-boson is massless only at tree level and (see
Fig.~1) acquires the {\it finite} mass $m_W^2=g^2\sqrt{v^2\alpha}/(16\pi)$ at
one loop,

\bigskip\medskip 
\vbox{
\be{Wmass}
\hskip 4truecm {=\frac{g^2\sqrt{v^2\alpha}}{16\pi}\delta_\mn \delta^{ab}\ .}
\ee 
\vskip-2.0truecm
\hskip .5truecm\psfig{figure=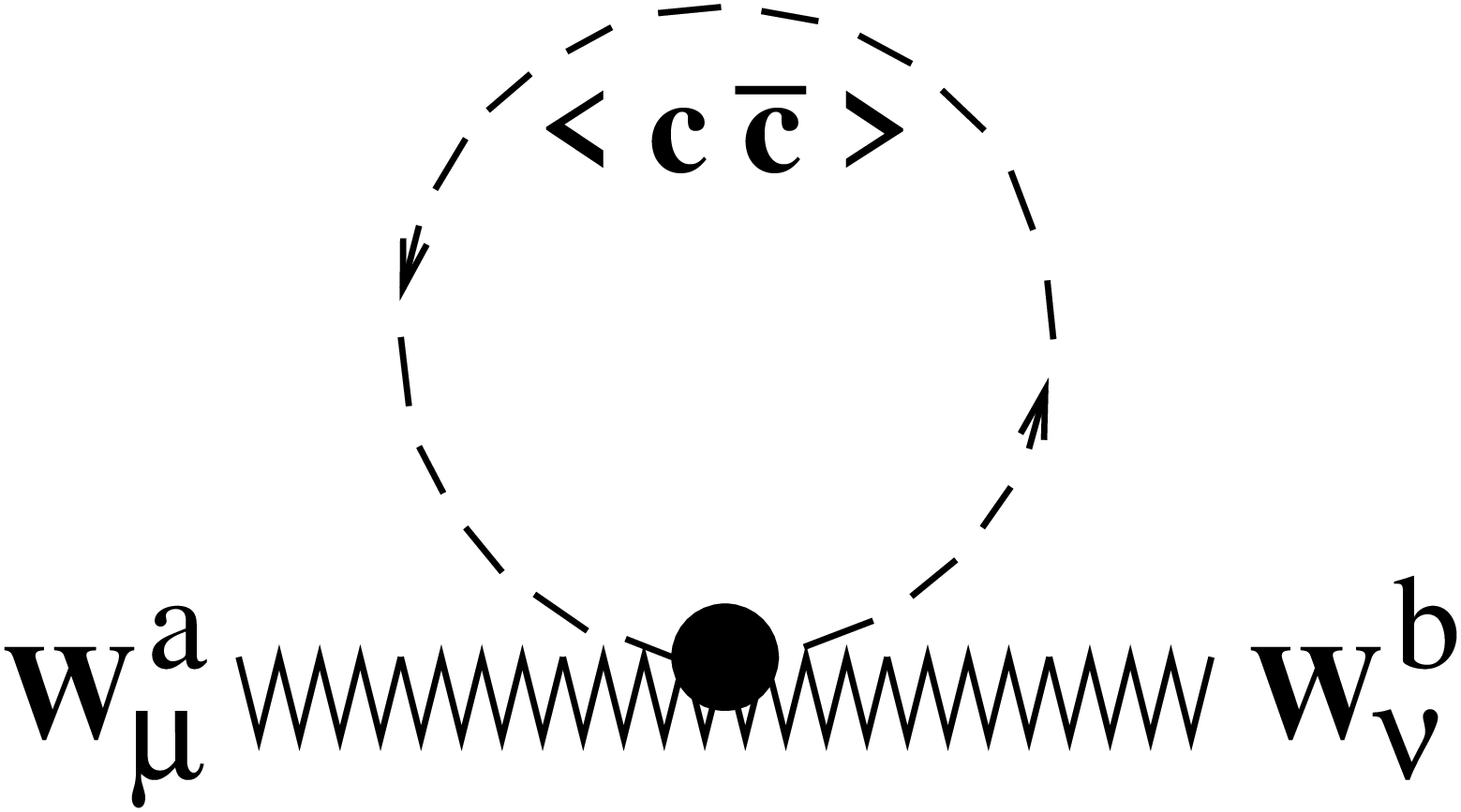,height=2.0truecm}
\\*
{\small\noindent Fig.~1. The finite one-loop contribution to the
$W$ mass.}}
\vskip 5pt
Technically, the one-loop contribution is finite because the integral 
in\equ{Wmass} involves only the $\delta^{ab}$-part
of the ghost propagator\equ{ghost}.
Since $p^2/(p^4+\alpha v^2)=-\alpha v^2/(p^2(p^4+\alpha v^2))+1/p^2$, the 
$v$-dependence of the loop integral is IR- {\it and}
UV-finite. The quadratic UV-divergence of the $1/p^2$ 
subtraction at $v=0$ is canceled by the other,
$v$-independent, quadratically divergent one-loop contributions -- (in
dimensional regularization this scale-invariant integral 
vanishes by itself). $m_W^2$ furthermore is {\it positive}
due to the overall minus sign of the {\it ghost} loop. The sign of $m_W^2$ is
crucial, for it indicates that the model is {\it
stable} and (as far as the loop expansion is concerned)
does not develop tachyonic poles at Euclidean $p^2$ for $v\neq 0$. 
Conceptually, the local mass term proportional to
$\delta_\mn\delta^{ab}$ is finite due to the
BRST symmetry\equ{brs}, which excludes a mass counter-term. 
The latter  argument implies that contributions to  $m^2_W$ are finite to all
orders of the loop expansion. 

\hskip 1.5truecm\psfig{figure=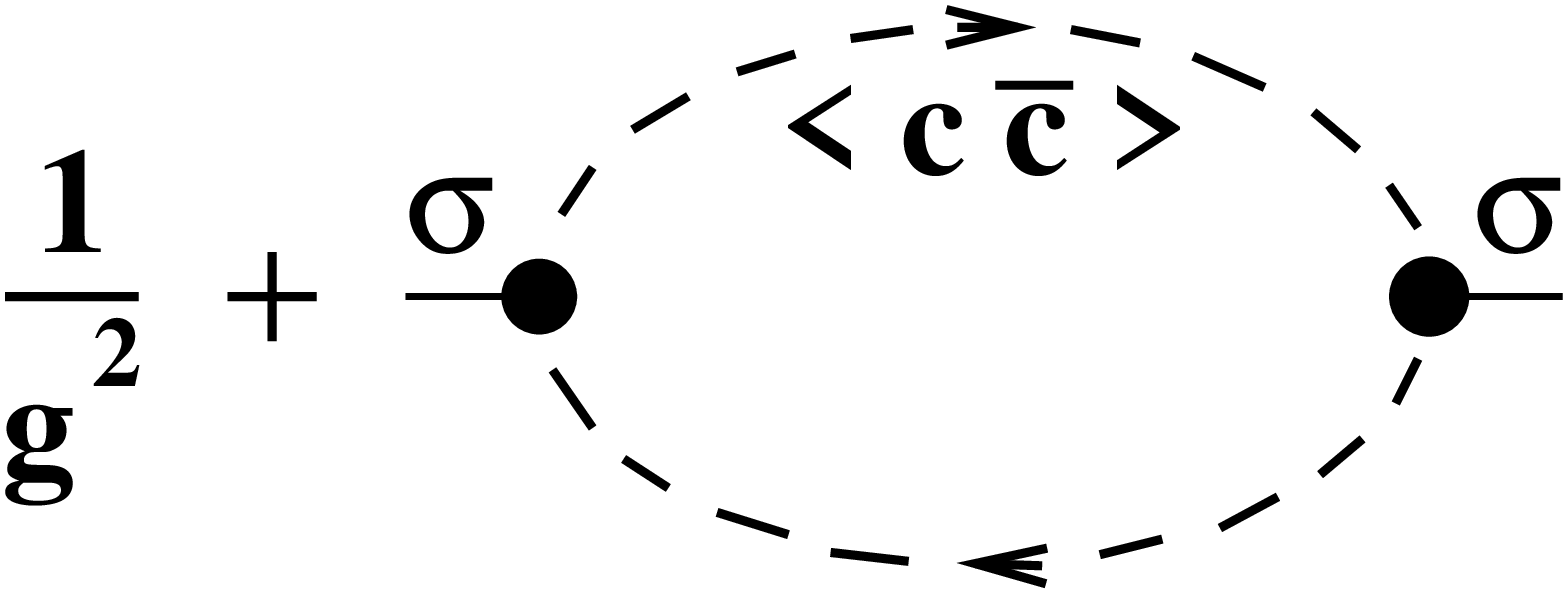,height=1.5truecm}
\\*
{\small\noindent Fig.~2. $\Gamma_{\sigma\sigma}(v,p^2)$ to
order $g^0$.}   
\vskip 5pt
If the model is stable at $v\neq 0$, the 1PI 2-point
function $\Gamma_{\sigma\sigma}(v, p^2)$ of the scalar must not vanish
at Euclidean $p^2$ either. To order $g^0$, $\Gamma_{\sigma\sigma}(p^2)$ is
given by the $1/g^2$ term that arises from \equ{Stratanovic} upon
substitution of\equ{decomp} and the one-(ghost)-loop contribution
shown in Fig.2. Since a non-trivial solution to the gap
equation\equ{gap} relates $1/g^2$ to a loop integral of zeroth order
in the coupling, we may use 
\equ{gap} to lowest order to  obtain a  ``tree-level''
expression for $\Gamma_{\sigma\sigma}(v,p^2)$ of order $g^0$. 
Evaluating the loop integrals, one obtains the real, positive and
monotonic function 
\ba{ss}
\Gamma_{\sigma\sigma}(x:=\frac{\sqrt{\alpha v^2}}{p^2})&=&\left\{\frac{
-1+2\sqrt{1-4ix}\,{\rm acoth}(\sqrt{1-4ix})}{32\pi^2\alpha^{-1}}\right\}\nonumber\\
&& + \left\{x\rightarrow -x\right\} \ .
\ea
%
$\Gamma_{\sigma\sigma}(p^2\geq 0)\ge\alpha/(16\pi^2)$ to order
$g^0$  establishes the 
perturbative stability of a non-trivial solution to\equ{gap} and
the fact that this solution is a minimum of the
effective potential.

An expansion about a  solution $v\neq 0$
to the gap equation thus has lowest order propagators that are regular
at Euclidean momenta for all the elementary 
fields  except the photon $A_\m$ (if all the matter fields are
massive).  The polarization of the photon
vanishes at $p^2=0$ due to the $U(1)$-symmetry -- regardless of the
value of $v$.  Taking into account that the massless Goldstone quartet
associated with this symmetry breaking decouples from physical
quantities, the situation for $v\neq 0$ is thus rather 
similar to QED with an unorthodox massive matter content (extending
the notion of ``massive matter'' to include ghosts and other 
unphysical fields). But the perturbative expansion of the free energy of
QED does not suffer from IR-divergences if all the
matter fields are massive\cite{Ka89}. Since all the
loop integrals of the  sceleton expansion of the free energy are
infrared finite at a non-trivial solution to\equ{gap}, 
a perturbative evaluation of the free energy of this
asymptotically free model is feasible and becomes accurate
at high temperatures.

To complete the argument, we have to solve\equ{gap} for small coupling.  
To lowest order in the loop expansion, the relation between the renormalized 
couplings $g,\alpha$, the renormalization point $\m$ and an expectation value
$v\neq 0$ implied by \equ{gap} is
\be{solg}
\ln{\frac{\alpha v^2}{\m^4}}=-\frac{16\pi^2}{\alpha g^2}+2+O(g^2)\ .
\ee
The anomalous dimension $\gamma_v$ of the expectation value is
simultaneously found to be
\be{gammav}
\gamma_v=-\frac{d\ln Z_v}{d\ln \m}=\frac{g^2}{16\pi^2}
(2\alpha -\beta_0) +O(g^4)\ , 
\ee
where $\beta_0$ is the lowest order coefficient of the
$\beta$-function of this model ($\beta_0=(22-2 n_f)/3$ with $n_f$
quark flavors in the fundamental representation as matter).

Using the relation between $\m$, $g^2$ and the asymptotic scale
parameter $\LQCD$, we may rewrite \equ{solg} as
\be{solQCD}
\ln{\frac{\alpha v^2}{\LQCD^4}}=\frac{16\pi^2}{g^2}\left(\frac{2}{\beta_0}-
\frac{1}{\alpha}\right)+2+O(\ln{g}, g^2)\ ,
\ee
Apart from an anomalous dimension, the non-trivial solution $v$ at
sufficiently small coupling is thus
proportional to the physical scale $\LQCD^2$ in the {\it particular}
gauge  $\alpha=\beta_0/2$. 
The anomalous dimension $\gamma_v$ in\equ{gammav} is furthermore of
order $g^4$ at $\alpha=\beta_0/2$.  The terms of order $\ln{g}$
in\equ{solQCD} thus also vanish in this particular gauge and higher order
corrections to the asymptotic value of 
$v$ at small $g^2$ are analytic in $g^2$. 
In the gauge $\alpha=\beta_0/2$, one can expand the model about
\be{start}
v^2=\frac{2}{\beta_0} e^2\LQCD^4 (1+O(g^2))
\ee
and determine the $O(g^2)$ corrections in \equ{start} order
by order in the loop expansion of the gap equation\equ{gap}. 
At $\alpha=\beta_0/2$ the lowest order solution\equ{start} remains
accurate to order $g^2$ at any finite order of the loop expansion. 
This {\it does not} imply that other gauges are any less physical, 
but it does single out $\alpha=\beta_0/2$ as the gauge in which a {\it
perturbative} evaluation of the gap equation\equ{gap} is consistent
at sufficiently small values of $g^2$. (In QED the hydrogen spectrum
to lowest order is most readily obtained in Coulomb gauge, although it 
evidently does not depend on the chosen gauge. In the present case
asymptotic freedom allows us to determine an optimal gauge for solving
the gap equation.)   

At the one-loop level,  \equ{gap} has a unique
non-trivial solution in any gauge $\alpha\neq 0$ and we 
know from $\Gamma_{\sigma\sigma}$ of\equ{ss}  that it corresponds to a
minimum of the one-loop action. In the limit $\alpha\rightarrow 0$ at
finite coupling, the non-trivial one-loop solution \equ{solg} coincides with
the trivial one. On the other hand, some of the couplings in the non-linear
gauge-fixing ${\cal L}_{\rm AG}$ become  
large in this limit, invalidating the perturbative analysis. 

To gain some insight into the highly singular behavior of the model
when $\alpha\sim 0$, I calculated the divergent part of the $W$
self-energy to one loop.
The corresponding anomalous dimensions $\gamma_W$ and
$\gamma_\alpha$ of the vector boson and the gauge parameter are 
\ba{anom}
\gamma_W&=&-\frac{d\ln Z_W}{d\ln
\mu}=\frac{g^2}{8\pi^2}\left(\beta_0-\frac{9}{2}- \frac{\alpha}{2}-\xi
\right)+O(g^4)\ ,\nonumber\\
\gamma_\alpha &=&\frac{d\ln{\alpha}}{d\ln \m} = 
-\frac{g^2}{8\pi^2}\left(\frac{3}{\alpha}+6-\beta_0 +
\alpha\right)+O(g^4).
\ea
Gauge dependent interaction terms proportional to
$g/\alpha$ at one loop thus lead to a term of order $g^2/\alpha^2$ in the
longitudinal part of the $W$ self-energy only. 
The transverse part of the $W$~self-energy is
regular in the limit $\alpha\rightarrow 0$. Taking $\alpha$ to vanish
thus is rather tricky:
\equ{anom} implies that the longitudinal part of the $W$-propagator at
one loop is proportional to $3
g^2 p^2\ln(p^2)$ at large momenta and no longer vanishes in this limit. 
Higher order loop corrections similarly
contribute to the longitudinal propagator as $\alpha\rightarrow 0$. 
$\gamma_\alpha$ does not depend on the gauge parameter $\xi$ at one loop, 
due to an Abelian Ward identity that also gives the QED-like
relation\cite{Re98} $Z_A=Z_g^{-2}=Z_\xi$  between the renormalization
constants of the photon,  of the coupling $g$
and of the gauge parameter $\xi$. 

The anomalous dimension of the gauge
parameter $\alpha$ at sufficiently small $g^2$ 
is negative for positive values of $\alpha$ when $\beta_0<6+2\sqrt{3}$. 
With $\gamma_\alpha<0$, 
the effective gauge parameter tends to decrease at higher
renormalization scales $\m$ and direct integration of \equ{anom} gives a
vanishing $\alpha$  at a {\it finite} value of the coupling $g^2$.  
As already noted above, the loop expansion, however, is valid only
if $g^2\ll 1$ and $g^2/\alpha\ll 1$. But \equ{anom} does show 
that there is no {\it finite} UV fixed point for the gauge
parameter and that $\alpha$ effectively  vanishes at least as
fast as $g^2$ as $\m\rightarrow \infty$ for any  gauge at finite $g^2$. 
\equ{start} nevertheless is the asymptotic solution to \equ{gap} in the
sense that it is  valid at arbitrary small coupling if one chooses 
the gauge at that coupling to be $\alpha(g)=\beta_0/2$. 
This is compatible with the asymptotic vanishing of the
effective gauge parameter only if higher order 
corrections lead to an anomalous dimension $\gamma_v$ that
effectively remains of order $g^4$ even 
as $g,\alpha(g)\rightarrow 0$. Since the gauge sector becomes strongly
coupled when $g^2/\alpha(g)$ is $O(1)$,
and the loop expansion
does not give the correct behavior of \equ{gammav} in this limit, 
this is at least conceivable.

Let me finally say that the non-trivial
solution to the gap equation apparently persists to arbitrarily high
temperatures. The (unique) non-trivial solution at $T=0$ is a
consequence of the scale anomaly\cite{Sc98} and the Goldstone quartet
of the spontaneously broken SL(2,R) symmetry does not contribute to
the free energy. The renormalization point dependence of\equ{solg}
and the associated UV-divergence of the loop  integral are an
indication of this. The character of the solution to\equ{gap} does,
however, change dramatically with temperature\cite{new}. At low 
temperatures $v(T)$ deviates only marginally from\equ{start}, whereas at 
high temperatures $v(T\sim\infty)\propto T^2/\ln^2{(T/\Lambda)}$.    

I would like to thank D.~Kabat, D.~Zwanziger and R.~Alkofer for 
suggestions,  L.~Spruch for his continuing support, and L.~Baulieu for
 encouragement. 

\end{document}